\documentclass[aip, reprint, citeautoscript]{revtex4-2}
\usepackage{graphicx}
\usepackage{amsmath, amssymb, amsfonts}
\usepackage{bm}
\usepackage[dvipsnames]{xcolor}

\begin{document}

\title{Converging on bound states in coupled-channel calculations}

\author{C. Ruth Le Sueur}
\affiliation{Joint Quantum Centre (JQC) Durham-Newcastle, Department of Chemistry, Durham University, South Road, Durham, DH1 3LE, United Kingdom.}

\author{James F. E. Croft}
\affiliation{Joint Quantum Centre (JQC) Durham-Newcastle, Department of Chemistry, Durham University, South Road, Durham, DH1 3LE, United Kingdom.}

\author{Jeremy M. Hutson}
{\email{j.m.hutson@durham.ac.uk} \affiliation{Joint Quantum Centre (JQC) Durham-Newcastle, Department of Chemistry, Durham University, South Road, Durham, DH1 3LE, United Kingdom.}

\date{\today}

\begin{abstract}
{\bf Abstract:} We develop a robust algorithm for locating bound states in coupled-channel calculations. Bound states exist at energies where an individual eigenvalue of a log-derivative or ratio matching matrix passes through zero. We describe an algorithm to identify the required eigenvalue of the matching matrix over the full range of energy where it exists. This allows much simpler programming than previous methods. We also consider the choice of the matching distance $R_\textrm{match}$, where the matching matrix is defined; coupled-channel methods are most efficient if $R_\textrm{match}$ is chosen to be in the classically allowed region for all channels that support bound states of interest, but not very close to a node in the wavefunction.
\end{abstract}

\maketitle

\section{Introduction}
The quantum bound-state problem is ubiquitous in physics and chemistry, in problems ranging from particle physics to chemical spectroscopy. There are many approaches to solving the multidimensional Schr\"odinger equation. Some methods use basis sets for all coordinates \cite{Bacic:1989, Tennyson:1991, Bowman:variational:2008}. At the other extreme, there are methods that use multidimensional grids \cite{Truhlar:1976, Ram-Mohan:1990}.

An alternative approach is the {\it coupled-channel} method \cite{Gordon:1969, Johnson:1978, Danby:method:1983, Hutson:CPC:1994, bound+field:2019}. This handles one coordinate $R$ by direct numerical propagation on a grid, and all the rest using a basis set. This avoids the use of a basis set for the $R$ coordinate, for which basis-set convergence is often poor, particularly when there is wide-amplitude motion in $R$.

Coupled-channel bound-state methods have been used extensively in the study of weakly bound systems such as Van der Waals complexes \cite{Dunker:1976, Danby:H2-H2:1983, Hutson:spher:1994} and in fitting potential-energy surfaces \cite{H92ArHF, H92ArHCl, Dubernet:1993, Carrington:hear:1995, H96ArCO2fit} to spectroscopic measurements \cite{Robinson:ArHCl-Pi:1987, Robinson:ArHCl-Sigma:1988, Nesbitt:ArHCl:FDCS:1988, Lovejoy:ArHCl:1988, Lovejoy:ArHF:1989, Ber92} They have also been used to study geometric-phase effects \cite{Kendrick:HO2:1997} and for the near-threshold bound states of ultracold molecules \cite{Hutson:Cs2:2008, Brue:AlkYb:2013, Karman:dipole:2018, Frye:high-spin:2020, Mukherjee:alkali:2024, Mukherjee:fl-bound:2026}, including associated work to fit interaction potentials \cite{Takekoshi:RbCs:2012, Berninger:Cs2:2013, Brookes:2022, Beulenkamp:2025}. They have particular advantages for states that lie close to dissociation thresholds, even when there are large numbers of much deeper states \cite{Croft:Rb+KRb:2026}. They allow ready incorporation of fine \cite{Mukherjee:RbYb:2022} and hyperfine \cite{Hutson:Cs2:2008} structure and of external electric \cite{Mukherjee:alkali:2024}, magnetic \cite{Gonzalez-Martinez:2007} and radiofrequency \cite{Owens:rf:2016} fields. They can locate bound states as a function of parameters other than energy, such as applied fields \cite{Takekoshi:RbCs:2012} or potential scaling factors \cite{Green:chaos:2016}. They also provide an integrated treatment of bound states and scattering: the same channel basis sets and propagation grids can be used for both, which is very valuable when searching for narrow Feshbach resonances \cite{Frye:high-spin:2020}.

Coupled-channel methods nevertheless present some difficulties in identifying the number of bound states in a particular range of energy and in converging efficiently on them. The object of the paper is to describe and document the best solutions we have found to these problems. They include an improved method of calculating the \emph{multichannel node count} and a simplified method for efficient convergence on bound-state energies.

\section{Methods}

\subsection{Coupled-channel equations}

The Hamiltonian of a system of two structured particles that interact in 3 dimensions is commonly of the form
\begin{equation}
	{\cal H}=-\frac{\hbar^2}{2\mu}\frac{1}{R}\frac{d^2}{dR^2} R + H_{\rm int}(\xi) + V(R,\xi),
\label{eq:Ham}
\end{equation}
where $R$ is a radial coordinate and $\xi$ represents all the other coordinates in the system. $H_{\rm int}$ represents the internal Hamiltonians of the two particles, and depends on $\xi$ but not $R$, and $V(R,\xi)$ is an operator that includes both the interaction potential and the operator for relative rotation.

The essence of coupled-channel methods is to use a basis-set expansion for all the coordinates \emph{except} $R$, and then to handle the dependence on $R$ by direct numerical propagation on a grid. The total wavefunction $\Psi$ is expressed as an $R$-dependent linear combination of orthonormal basis functions $\Phi_j(\xi)$,
\begin{equation}
	\Psi=\frac{1}{R}\sum_j \Phi_j(\xi)\psi_j(R).
\label{eq:expand}
\end{equation}
The prefactor $R^{-1}$ serves to simplify the form of the radial kinetic energy operator. The radial channel functions $\psi_j(R)$ gives the $R$-dependence of the wavefunction for channel $j$.  Substituting this expansion into the Schr\"odinger equation and projecting onto each basis function $\Phi_i(\xi)$ in turn gives a set of coupled differential equations that can be expressed in matrix-vector form,
\begin{equation}
	\frac{d^2\boldsymbol{\psi}}{dR^2}=\left[ {\bf W}(R) - {\cal E}{\bf I} \right]\boldsymbol{\psi}(R),
\label{eq:coupled}
\end{equation}
where ${\cal E}=2\mu E/\hbar^2$ is a reduced energy.
If $N$ channel basis functions $\Phi_j(\xi)$} are included in the expansion, $\boldsymbol\psi(R)$ is a column vector of order $N$, ${\bf I}$ is the $N\times N$ unit matrix and ${\bf W}(R)$ is an $N\times N$ matrix with elements
\begin{equation}
	W_{ij}(R)=\frac{2\mu}{\hbar^2}\int \Phi_i^*(\xi)\left[H_{\rm int}(\xi)+V(R,\xi)\right]\Phi_j(\xi)\,d\xi.
\end{equation}
The different channels are coupled by the off-diagonal elements of ${\bf W}$. It should be emphasized that the number of channels $N$ is usually much smaller than the total number of basis functions needed in methods that use basis sets in all coordinates \cite{Bacic:1989, Tennyson:1991, Bowman:variational:2008}.

Very similar matrix equations of the form (\ref{eq:coupled}) are obtained in other contexts. These include two particles interacting in 1 or 2 dimensions, pairs of atoms in separated optical tweezers \cite{Bird:mergo:2023} and systems of 3 or more particles treated in hyperspherical coordinates \cite{PACK:1987}.

In practice, propagating the wavefunction matrix itself may cause numerical instabilities. Instead, it is usual to propagate the log-derivative matrix ${\bf Y}={\bf\Psi}^{-1} {\bf \Psi}'$, where ${\bf \Psi'}=d{\bf\Psi}/dR$~\cite{Johnson:1973, Alexander:1984, Manolopoulos:1986, MG:symplectic:1995}. Alternatively, it is possible to propagate a ratio of wavefunction matrices at successive propagation steps, usually by the renormalized Numerov method~\cite{Johnson:1978}. We focus here on log-derivative methods, as implemented in the general-purpose packages BOUND and FIELD~\cite{bound+field:2019, mbf-github:2025}, but describe the extension to the renormalized Numerov method briefly in Sec.\ \ref{sec:renorm}.

\subsection{Bound-state calculations}

In coupled-channel bound-state calculations, the coupled equations are solved subject to boundary conditions at both short and long range. These most commonly require that $\psi_j(R) \rightarrow 0$ as $R\rightarrow0$ and $\infty$, but are different in some cases. In general there are $N$ linearly independent solution vectors $\boldsymbol{\psi}_k$ that satisfy the boundary conditions at \emph{one} end of the range at any energy $E$, but none that satisfy them at \emph{both} ends of the range unless $E=E_m$, where $E_m$ is one of the allowed bound-state energies. The central problem of bound-state calculations is to find the energies $E_m$ and the corresponding solution vectors $\boldsymbol{\psi}^\textrm{b}_m(R)$.

The quantized solution vectors $\boldsymbol{\psi}^\textrm{b}_m(R)$ may be expanded in terms of the $N$ linearly independent solution vectors $\boldsymbol{\psi}_k$ at energy $E_m$. However, the particular linear combination of $\boldsymbol{\psi}_k$ that is required is not known until \emph{after} an allowed energy $E_m$ is found. Before this, it is useful to solve Eq.\ \ref{eq:coupled} at trial energy $E$ to find an $N\times N$  wavefunction \emph{matrix} ${\bf\Psi}$ that is formed by stacking the $N$ independent solution vectors side-by-side~\cite{Gordon:1969}.

The usual approach is to propagate one matrix solution of Eq.\ \ref{eq:coupled} outwards from $R=0$ (or from $R_\textrm{min}$, deep in the classically forbidden region at short range) and another inwards from $R_\textrm{max}$, deep in the classically forbidden region at long range. The two solutions are matched at a distance $R_\textrm{match}$ in the classically allowed region.

A bound-state wavefunction must be continuous and differentiable. This requires that the wavefunction $\boldsymbol{\psi}$ and its radial derivative $\boldsymbol{\psi}'$ from inwards and outwards propagations must be equal at $R_\textrm{match}$.
This requirement is particularly simply expressed in terms of the log-derivative matrix, since $\boldsymbol{\psi}'={\bf Y}\boldsymbol{\psi}$,
\begin{align}
\boldsymbol{\psi}'(R_\textrm{match}) &= {\bf Y}^+_\textrm{i}(R_\textrm{match}) \boldsymbol{\psi}(R_\textrm{match}) \nonumber\\
&= {\bf Y}^-_\textrm{o}(R_\textrm{match}) \boldsymbol{\psi}(R_\textrm{match})
\end{align}
where $+$ and $-$ indicate outwards and inwards propagations, respectively, and i and o indicate the regions inside and outside $R_\textrm{match}$.
Equivalently,
\begin{equation}
		{\bf Y}_\textrm{match} \boldsymbol{\psi}(R_\textrm{match}) = 0,
\label{eq:Ymatch}
\end{equation}
where ${\bf Y}_\textrm{match} = {\bf Y}^+_\textrm{i}(R_\textrm{match}) - {\bf Y}^-_\textrm{o}(R_\textrm{match})$.

Equation \ref{eq:Ymatch} has a non-trivial solution only if the determinant $|{\bf Y}_\textrm{match}|$ is zero. Early approaches to locating bound states using coupled-channel calculations~\cite{Gordon:1969, Johnson:1978, Danby:method:1983, Hutson:bound:1984} relied on searching numerically for zeroes in this determinant (or a closely related one from a different propagator) as a function of energy. However, the determinant often has properties that are numerically undesirable, such as dipping briefly below zero and back again as a function of energy. Hutson~\cite{Hutson:CPC:1994} pointed out that the individual eigenvalues $\epsilon_i$ of ${\bf Y}_\textrm{match}$ have much more desirable properties. In particular, they decrease monotonically as a function of energy, except where they pass through poles \cite{Manolopoulos:PhD:1988}. Moreover, Eq.\ \ref{eq:Ymatch} indicates that $\boldsymbol{\psi}(R_\textrm{match})$ is an eigenfunction of ${\bf Y}_\textrm{match}$ with eigenvalue zero. Complete multichannel wavefunctions may be obtained by back-substituting into the log-derivative propagation equations \cite{Thornley:1994}, starting from this eigenvector at $R_\textrm{match}$.

\section{Results and Discussion}

\subsection{Behavior of eigenvalues of matching matrix}

\begin{figure}
\includegraphics[width=0.49\textwidth]{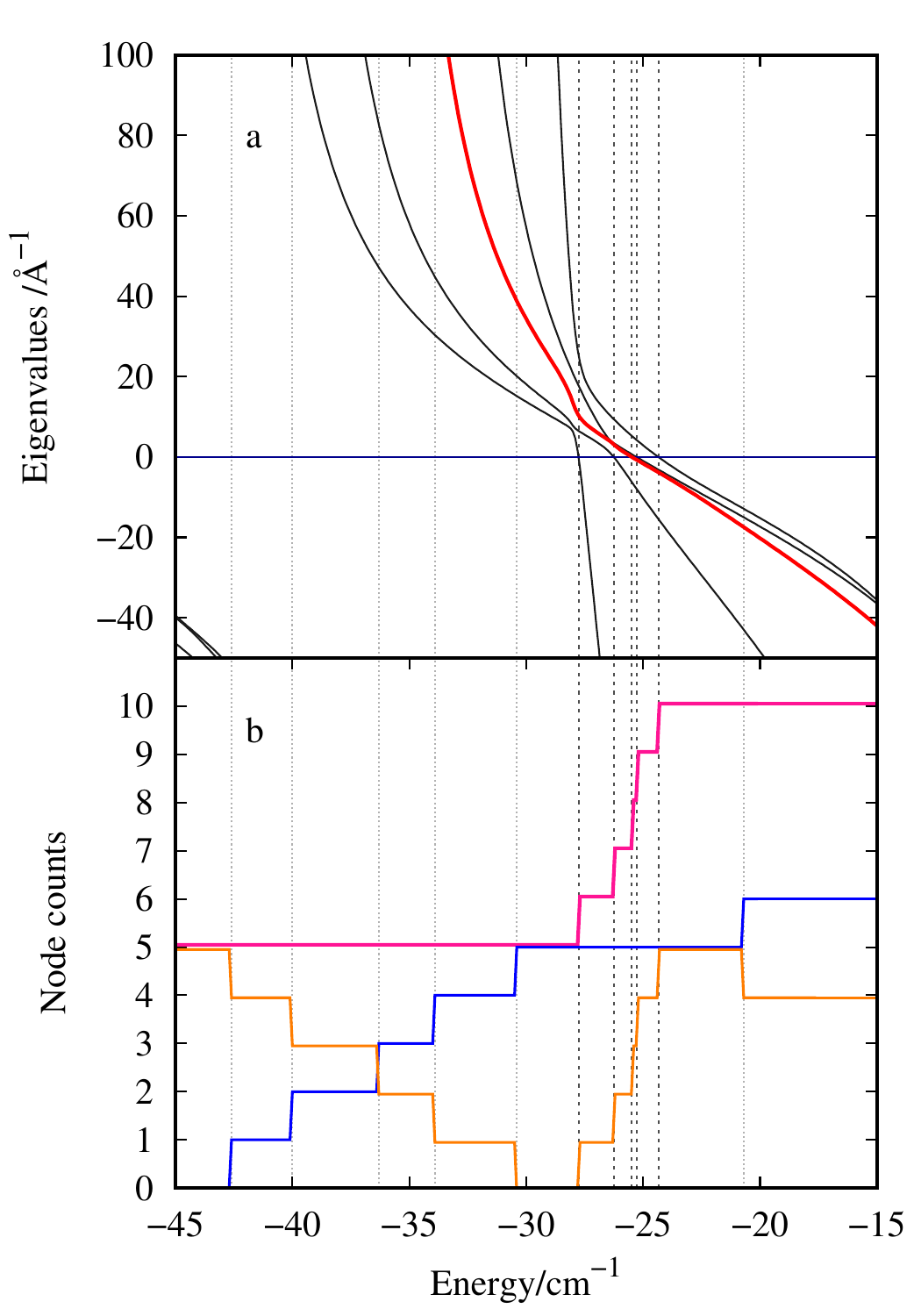}
\caption{(a) Eigenvalues of the 5-channel matching matrix for the E states of Ar-CH$_4$ with $J=2$ and $j_\textrm{max}=2$, with $R_\textrm{match}$ chosen to be at the potential minimum. The eigenvalue corresponding to a representative state is highlighted in red.
(b) Node count $n(E)$ (pink) and contributions to it: $n_\textrm{sum}(E)$ (blue) from propagation and $n_\textrm{match}(E)$ (orange) from the matching matrix.
Bound-state energies are shown with vertical dashed lines and pole positions with dotted lines.}
\label{fig:eig-and-nodes}
\end{figure}

Figure \ref{fig:eig-and-nodes}(a) shows an example of the behavior of the eigenvalues of ${\bf Y}_\textrm{match}$, calculated for the E states of Ar-CH$_4$ \cite{Hutson:spher:1994} using the interaction potential of ref.\ \onlinecite{Buck:1983}. This simple example is for total angular momentum $J=2$ and includes only the CH$_4$ states with $j=2$ (which is the lowest allowed for E symmetry). $R_\textrm{match}$ is chosen here to be at the potential minimum. There are 5 coupled channels, and 5 states that lie between $-28$ and $-24$ cm$^{-1}$, measured with respect to the dissociation energy for $j=0$.
Since the basis set cannot be factorized into blocks of distinct symmetry, the eigenvalues do not cross one another as a function of energy, and instead show avoided crossings. The line highlighted in red corresponds to the eigenvalue for one state as it passes through avoided crossings between its poles.

\subsection{Multichannel node count}

To converge reliably on all the allowed energies $E_m$ in a particular range, it is necessary to enumerate them. This may be done using the multichannel node count $n(E)$ \cite{Calvert:1969}. This is defined at all energies $E$ as the number of \emph{focal points} where $|\boldsymbol{\Psi}(R)|=0$. Johnson \cite{Johnson:1978} gave an intuitive demonstration that $n(E)$ is equal to the number of bound states that lie below energy $E$; in that sense it implicitly includes nodes in all coordinates. It increases monotonically with energy, and increases by one at the energy of each bound state.

The contribution to $n(E)$ from a single propagation step is easily evaluated as a by-product of propagation. In log-derivative propagation, a zero in $|\boldsymbol{\Psi}(R)|$ corresponds to a pole in $|\boldsymbol{Y}(R)|$; such poles manifest themselves as negative eigenvalues of a real symmetric matrix $\boldsymbol{Z}$ that is inverted during propagation; further details are given in ref.\ \onlinecite{Hutson:CPC:1994}. This requires very little extra computation, because the number of negative eigenvalues of a real symmetric matrix can be calculated as a by-product of its inversion \cite{Bunch:1977}. The contribution to $n(E)$ from a propagation segment (containing many steps) is the accumulated number of negative eigenvalues of $\boldsymbol{Z}$ that are encountered across the segment.

In Johnson's original bound-state method \cite{Johnson:1978}, $n(E)$ is evaluated by propagating across the entire range of $R$ in a single direction,
\begin{equation}
n = n^+_\textrm{i}+n^+_\textrm{o} = n^-_\textrm{i}+n^-_\textrm{o},
\label{eq:n_Johnson}
\end{equation}
where $n^\pm_\textrm{i}$ and $n^\pm_\textrm{o}$ are contributions from individual propagation segments; $+$ and $-$ indicate outwards and inwards propagations and i and o indicate the regions inside and outside $R_\textrm{match}$. Evaluating \emph{both} ${\bf Y}_\textrm{match}$ and $n$ thus requires three propagation segments, with propagations in both directions across either the inner region or the outer region.

A more efficient approach is based on only two propagations, carried out from small and large $R$ to the matching point at $R_\textrm{match}$. Manolopoulos \cite{Manolopoulos:PhD:1988} showed that the node count is then
\begin{equation}
n = n_\textrm{sum} + n_\textrm{match}
\label{eq:n_match}
\end{equation}
where $n_\textrm{sum}=n^+_\textrm{i}+n^-_\textrm{o}$ and $n_\textrm{match}$ is the number of negative eigenvalues of ${\bf Y}_\textrm{match}$. This version of the node count has been implemented in recent versions of the BOUND package \cite{bound+field:2019, mbf-github:2025}, but its behavior has not been described in the published literature.

Evaluating $n(E)$ from Eq.\ \ref{eq:n_match} instead of Eq.\ \ref{eq:n_Johnson} has a further important advantage when certain boundary conditions are used. For example, for bound states that lie just below a dissociation limit, it is necessary to propagate to very large $R$ before the wavefunctions approach close to zero. To ameliorate this, it is common to use a WKB (Wentzel-Kramers-Brillouin) boundary condition \cite{Hutson:CPC:1994} at a distance $R_\textrm{max}$ where the wavefunction is small but not close to zero. However, evaluating $n(E)$ from outwards propagation using Eq.\ \ref{eq:n_Johnson} omits focal points that occur outside $R_\textrm{max}$, and results in $n(E)$ changing at an energy slightly different from $E_m$. This problem is removed by using Eq.\ \ref{eq:n_match}.

Figure \ref{fig:eig-and-nodes}(b) shows the behavior of $n$, $n_\textrm{sum}$ and $n_\textrm{match}$ as functions of energy for the same case as Fig.\ \ref{fig:eig-and-nodes}(a), with vertical lines connecting related features. It may be seen that $n(E)$ increases by 1 each time an eigenvalue of ${\bf Y}_\textrm{match}$ passes through zero, so at each bound-state energy. Where an eigenvalue passes through a pole, $n_\textrm{sum}$ increases by 1 and $n_\textrm{match}$ decreases by 1~\cite{Manolopoulos:PhD:1988} so that $n(E)$ remains constant. Nevertheless, $n_\textrm{sum}$ and $n_\textrm{match}$ are individually important, as described below.

\subsection{Identifying the required eigenvalue of \protect{${\bf Y}_\textrm{match}$}}

A difficulty of converging on a zero in an individual eigenvalue of ${\bf Y}_\textrm{match}$ is how to identify numerically the specific eigenvalue to use and the region of energy over which it is monotonically decreasing. The general strategy is to use bisection of energy intervals until a suitable region of energies is identified, and then switch to a fast root-finding algorithm such as Brent's method \cite{Brent:1973}. Various approaches have been used to decide when to switch, mostly based on monitoring the smallest eigenvalue of each sign as a function of energy \cite{mbf-documentation:2025}. However, they are complicated to program and often identify the wrong eigenvalue at intermediate steps; this delays convergence, though it does not usually prevent it entirely. The present paper offers an improved approach.

We choose to label each bound state by the value of $n$ immediately above it, so that the lowest state is labeled $m=1$, as in the BOUND package \cite{bound+field:2019}. Each bound state can be associated with a single continuous eigenvalue of the matching matrix. However, as seen in Fig.\ \ref{fig:eig-and-nodes}, the \emph{index} of this eigenvalue (in a value-ordered array) changes whenever one of the other eigenvalues passes through a pole. The eigenvalue that passes through zero at the energy of bound state $m$ has index
\begin{equation}
i=m-n_\textrm{sum}(E).
\label{eq:index}
\end{equation}
This eigenvalue decreases monotonically between poles at $E_\textrm{low}$ and $E_\textrm{high}$, while its index decreases monotonically in a stepwise manner from $N$ to 1.  The eigenvalue that corresponds to bound state $m$ is available at energy $E$ only if
\begin{equation}
m-N \le n_\textrm{sum}(E)< m .
\label{eq:range}	
\end{equation}
If two or more states are symmetry-related and exactly degenerate, the corresponding eigenvalues are identical at all $E$ and Eqs.\ \ref{eq:index} and \ref{eq:range} remain valid.

\subsection{Convergence algorithm}

Our goal is usually to find the energies of all bound states that lie in the range $E_\mathrm{bottom}$ to $E_\mathrm{top}$. We first carry out log-derivative propagations at these two energies, giving node counts $n_\mathrm{bottom}$ and $n_\mathrm{top}$. This means that bound states $n_\mathrm{bottom}+1$ to $n_\mathrm{top}$ lie in the required range. We refer to these as target states.

\begin{figure}
\includegraphics[width=0.49\textwidth]{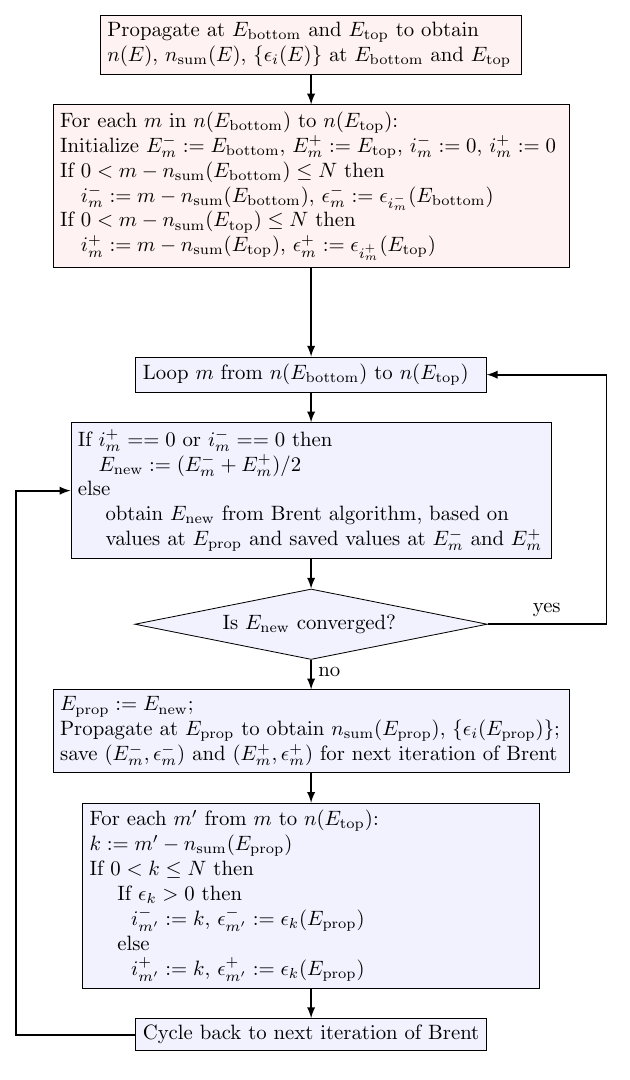}
\caption{Flow diagram of convergence algorithm, showing initialization (pink) and updating of the six arrays $E_m^-$, $E_m^+$, $i_m^-$, $i_m^-$, $\epsilon_{i_m^-}$ and $\epsilon_{i_m^+}$ (blue).}
\label{fig:flow}
\end{figure}

To avoid repeating calculations, we maintain six arrays of dimension $n_\mathrm{top}-n_\mathrm{bottom}+1$. For each target state $m$, these contain:
(1) The current lower bound $E_m^-$ on its energy, which is the highest energy so far with $n(E)<m$;
(2) The current upper bound $E_m^+$, which is the lowest energy so far with $n(E)\ge m$;
(3) and (4) The indices $i_m^\pm=m-n_\textrm{sum}(E_m^\pm)$ at the lower and upper bounds, set to 0 if outside the range 1 to $N$;
(5) and (6) The corresponding eigenvalues $\epsilon_{i_m^\pm}(E_m^\pm)$ of the matching matrix at the upper and lower bounds.
A diagram showing how these arrays are initialized and updated is given as Fig.\ \ref{fig:flow}.

A major advantages of the present method is its robustness. Both bisection and Brent's method ensure convergence so long as the root is bracketed. The monotonically decreasing eigenvalues of the matching matrix guarantee that the root is bracketed once a positive and a negative eigenvalue are identified. The method thus guarantees convergence, provided the two initial energies do bracket the bound state(s). It is particularly advantageous in cases with near-degenerate states, where previous algorithms (based on the smallest eigenvalue at each energy) often picked an incorrect eigenvalue during convergence.

\subsection{Effect of locally closed channels}

\begin{figure}
	\includegraphics[width=0.49\textwidth]{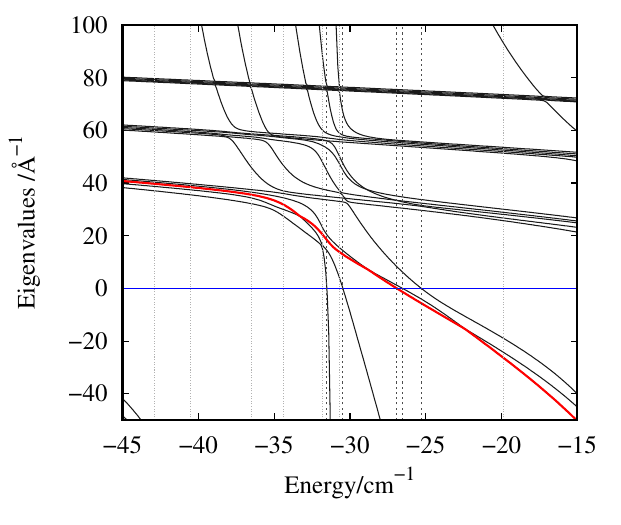}
	\caption{Eigenvalues of the matching matrix for the E states of Ar-CH$_4$ with $J=2$ and $j_\textrm{max}=6$. This case is the same as Fig.\ \ref{fig:eig-and-nodes}(a) but with additional locally closed channels. Bound-state energies are shown with vertical dashed lines and pole positions with dotted lines.}
	\label{fig:eig-closed}
\end{figure}

The eigenvalues of the matching matrix have more complicated behavior when there are locally closed channels at $R_\textrm{match}$. Figure \ref{fig:eig-closed} shows the eigenvalues for the same test case as in Fig.\ \ref{fig:eig-and-nodes}, but using $j_\textrm{max}=6$ to give a larger basis set. The closed channels introduce additional eigenvalues, with positive values that decrease only slowly with energy, as described in Sec.\ \ref{sec:Rmatch}. They have avoided crossings with the eigenvalues in Fig.\ \ref{fig:eig-and-nodes}. As a result, the eigenvalue for the highlighted state takes a more convoluted route through various avoided crossings, and in fact never reaches a pole at energies below its zero. Nevertheless, the condition of Eq.\ \ref{eq:range} is still able to identify the correct eigenvalue. The irregular behavior may slow convergence to some extent, but the algorithm is still guaranteed to converge. We have used it successfully in cases with very large numbers of closed channels.

\subsection{Choice of \protect{$R_\textrm{match}$}}
\label{sec:Rmatch}

The procedure described above is formally guaranteed to converge for any choice of $R_\textrm{match}$, but there are some choices that are inefficient and should be avoided.

\begin{figure}
\includegraphics[width=0.49\textwidth]{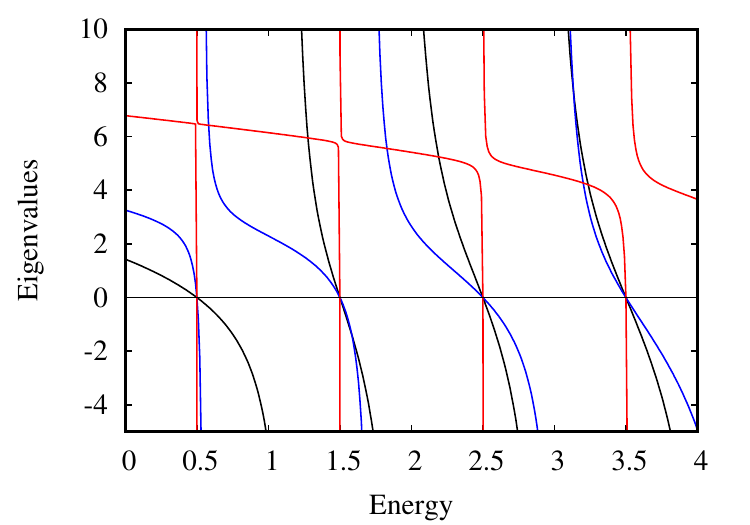}
\caption{Matching eigenvalue for a single-channel reduced harmonic oscillator with $R_\textrm{match}$ chosen in the classically allowed region (black), near the classical turning point for $v=1$ (blue), and well into the classically forbidden region (red).}
\label{fig:harm-class-forbid}
\end{figure}

First, $R_\textrm{match}$ should be chosen to be in (or near) the classically allowed region of every channel that supports bound states of interest. This is because wavefunctions decay exponentially in classically forbidden regions. Far from a classical turning point, a single-channel wavefunction in a locally closed channel may be approximated using the WKB (Wentzel-Kramers-Brillouin) approximation \cite{Child:1974}. As an example, consider the case where $R_\textrm{match}$ is inside the inner turning point $R_\textrm{in}$. An outgoing WKB wavefunction that satisfies a bound-state boundary condition grows exponentially in this region, with the functional form
\begin{equation}
\psi(R) \approx k^{-\frac{1}{2}}(R) \exp\left(-\int_R^{R_\textrm{in}} k(R')\,dR' \right),
\end{equation}
where $k(R)$ is the local wave vector, such that $\hbar^2 k^2(R) = V(R) - E$. Neglecting the slowly-varying prefactor of $k^{-1/2}$, this has log-derivative $k(R)$. The incoming function, at an energy $E$ that is not an eigenvalue of the Schr\"odinger equation, has both exponentially decaying and exponentially growing components as it is propagated towards $R=0$. Far in the classically forbidden region, the exponentially growing component dominates, and the log-derivative approaches $-k(R)$ except extremely close to the eigenvalue. The log-derivative matching function does still cross zero at the bound-state energy, where the exponentially increasing component vanishes exactly, but it is almost constant at $+2k(R)$ at all energies that are not very close to the eigenvalue. A numerical example of this behavior is shown in Fig.\ \ref{fig:harm-class-forbid} for the lowest few states of a reduced harmonic oscillator, with $R_\textrm{match}$ chosen at three different points: in the classically allowed region, near the classical turning point for $v=1$, and far into the classically forbidden region. When $R_\textrm{match}$ is in the classically forbidden region, the eigenvalue differs from its background value over only a small range of energies. Locating the zero of such a function typically requires many points. When $R_\textrm{match}$ is close to the classical turning point, the behavior is better but not optimum. It is best to place $R_\textrm{match}$ in the classically allowed region.

\begin{figure}
\includegraphics[width=0.49\textwidth]{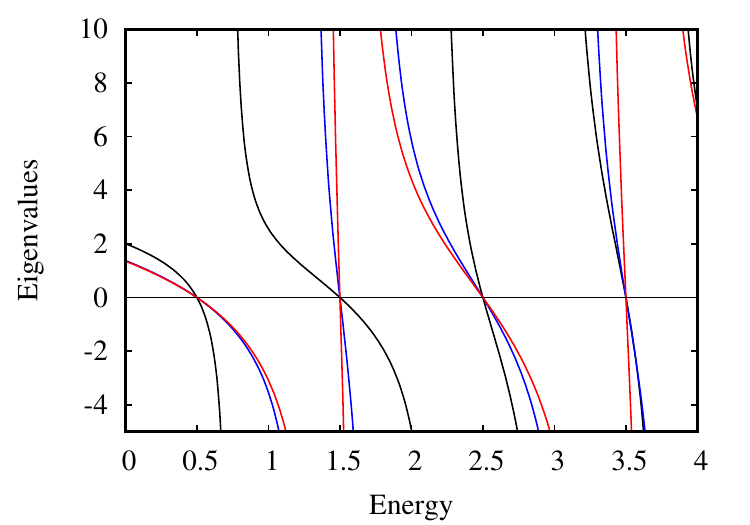}
\caption{Matching eigenvalue for a single-channel reduced harmonic oscillator with $R_\textrm{match}$ chosen far from the central node for odd $v$ (black) and at two values increasingly close to it (blue and red).}
\label{fig:harm-close-to-node}
\end{figure}

Another situation that can cause slow convergence is placing $R_\textrm{match}$ very close to a node in the wavefunction of a state of interest. In this case the log-derivative function still passes through zero at the corresponding eigenvalue of the Schr\"odinger equation, but it has poles very close to the zero on either side, which get closer as $R_\textrm{match}$ approaches the position of the node. This can once again require many bisections before a faster root-finding algorithm can be used. Figure \ref{fig:harm-close-to-node} shows an example of this behavior for a reduced harmonic oscillator, with $R_\textrm{match}$ chosen at two points close to the centre of the oscillator and one point far from it. For odd $v$ the wavefunction has a node exactly at the centre; when $R_\textrm{match}$ is placed close to that node, the zeroes in the eigenvalues for odd $v$ are placed between two closely spaced poles, which get closer together as $R_\textrm{match}$ approaches the node.

The combined effect is that it is usually best to place $R_\textrm{match}$ in the classically allowed region, close enough to the inner turning points of important channels to avoid any nodes in the wavefunction. In practice, however, problems due to nodes in wavefunctions are rare, and convergence is usually fast for any choice of $R_\textrm{match}$ in the classically allowed region. It is often sufficient to place $R_\textrm{match}$ close to the potential minimum, in a region where all important channels are locally open. In the rare cases that this causes problems because of the proximity of a node in a wavefunction, the problems can be circumvented with a slightly different $R_\textrm{match}$.

\subsection{Finding bound states as a function of a parameter other than energy}

There are some applications where it is necessary to locate bound states as a function of a parameter other than energy. In studies of ultracold atoms and molecules, for example, bound states are often required as a function of magnetic field at fixed energy \cite{Takekoshi:RbCs:2012, Berninger:Cs2:2013}. In studies of quantum chaos, bound states have been located as a function of potential scaling factor \cite{Green:chaos:2016, Frye:chaos:2016}. The method presented here can be used in such cases, as long as $n$ is a monotonically increasing or decreasing function of the varying quantity across the range specified. Within such a region, all the eigenvalues $\epsilon_i$ that cross zero do so in the same direction, and each eigenvalue crosses zero only once. If $n$ is not monotonic, it is possible to scan across the range to identify smaller regions where it is.

\subsection{Extension to renormalized Numerov propagation}
\label{sec:renorm}

The renormalized Numerov method is based on outwards propagation of the \emph{ratio} of wavefunction matrices at the two ends of each propagation step, ${\bf R}_n={\bf F}_{n+1} {\bf F}_n^{-1}$, in place of the log-derivative matrix ${\bf Y}$. Here ${\bf F}_n$ is a slightly modified version of $\boldsymbol{\Psi}_n$, transformed using the Numerov prescription as described by Johnson \cite{Johnson:1978}. The corresponding matrix for inward propagation is $\hat{\bf R}_n={\bf F}_{n-1} {\bf F}_n^{-1}$.
The matching matrix at $R_\textrm{match}$ (at point $m$) is then defined as
\begin{equation}
  {\bf R}_\textrm{match} = {\bf R}_m - \hat{\bf R}_{m+1}^{-1}.
\end{equation}
This has very similar properties to the log-derivative matching matrix, and all the methods described here can be applied to  ${\bf R}_\textrm{match}$ in place of ${\bf Y}_\textrm{match}$.

\section{Conclusions}

Coupled-channel methods for bound states rely on locating zeroes (roots) in either the determinant or a single eigenvalue of a matching matrix. Both these functions have zeroes and poles as a function of energy, but the individual eigenvalues have much better properties for root-finding because they decrease monotonically with energy. Previous algorithms for locating the zeroes have involved complicated programming and \emph{ad hoc} decisions based on the smallest eigenvalues at each energy. We have developed a more systematic algorithm that allows the eigenvalue corresponding to a particular bound state to be identified everywhere it exists. This allows much simpler programming and is usually more efficient.

We have also considered the choice of the distance $R_\textrm{match}$ at which the matching matrix is defined. We have shown that the eigenvalues have undesirable properties for root-finding if $R_\textrm{match}$ is far in the classically forbidden region for a channel that supports bound states of interest, or if it is very close to a node in the wavefunction of such a bound state. In practice, it is usually satisfactory to choose $R_\textrm{match}$ to be in the vicinity of the potential minimum.

We have implemented the algorithm developed here in the packages BOUND and FIELD \cite{bound+field:2019, mbf-github:2025}, and they will be included in a future public release of the programs.

\section*{Rights retention statement}

For the purpose of open access, the authors have applied a Creative Commons Attribution (CC BY) licence to any Author Accepted Manuscript version arising from this submission.

\section*{Data availability statement}

The data underlying this study are openly available from the Durham University Research Data Repository at DOI 10.15128/r1jd472w556.

\section*{Acknowledgement}
This work was supported by the U.K. Engineering and Physical Sciences Research Council (EPSRC) Grant Nos.\
EP/W00299X/1, 
EP/Z535898/1, 
UKRI1111,     
and UKRI2226. 

\section*{References}

\bibliography{all}

\begin{figure*}[p]
\includegraphics{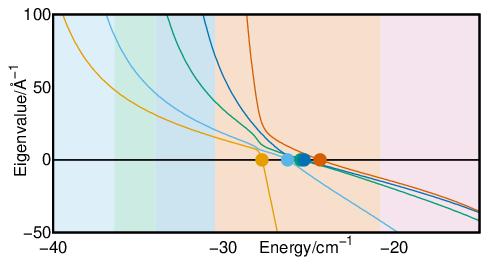}
\caption{Graphic for Table of Contents only}

\end{figure*}

\end{document}